\documentstyle[prl,aps]{revtex}
\begin{document}
\draft

\def\overlay#1#2{\setbox0=\hbox{#1}\setbox1=\hbox to \wd0{\hss #2\hss}#1%
\hskip
-2\wd0\copy1}
\twocolumn[
\hsize\textwidth\columnwidth\hsize\csname@twocolumnfalse\endcsname

\title{Relativistic Effects of Light in 
Moving Media with Extremely Low Group Velocity}
\author{U.\ Leonhardt$^{1,2}$ and P.\ Piwnicki$^1$}
\address{~$^1$Physics Department, Royal Institute of Technology (KTH),
Lindstedtsv\"agen 24, S-10044 Stockholm, Sweden}
\address{~$^2$School of Physics and Astronomy, University of St Andrews, 
North Haugh, St Andrews, Fife, KY16 9SS, Scotland}
\maketitle
\begin{abstract}
A moving dielectric medium acts as an effective gravitational field on
light. One can use media with extremely low group velocities 
{[}Lene Vestergaard Hau {\it et al.}, Nature {\bf 397}, 594 (1999){]}
to create dielectric analogs of astronomical effects on Earth.
In particular, a vortex flow imprints a long--ranging topological effect on
incident light and can behave like an optical black hole.
\end{abstract}
\date{today}
\pacs{04.20.-q, 42.50.Gy, 03.65.Bz, 03.75Fi}
\vskip2pc]
\narrowtext


According to general relativity, acceleration and gravitation are
equivalent in the absence of other forces. 
A freely falling test particle, seen in any local inertial system,
moves along a straight line. 
And yet, the inertial frames along the particle's path are 
non--trivially connected; space--time is curved such that the
trajectory is bent in general.
An analogous situation occurs when light propagates in a dielectric
medium \cite{Gordon,LP}.
Seen locally, light rays are straight lines in all volume elements of
the medium. 
Seen globally, the medium elements might move in different directions
and drag the light or the refractive index may vary such that light
rays are curved.
Seen in four--dimensional space--time, light follows a zero--geodesic
line with respect to a metric that comprises the medium's dielectric
properties \cite{Gordon,LP}.

Ordinary dielectrics require astronomical velocity gradients 
to establish some of the spectacular effects of general relativity, 
velocities that are comparable with the speed of light in the medium.
Recently, extraordinary dielectrics 
that are distinguished by an extremely low group velocity of light
have been made on Earth \cite{Lene}.
As we shall describe in this paper,
the reported experiment is sensitive enough to detect quantum vortices via
an optical Aharonov--Bohm effect.
Furthermore, a vortex may become an optical black hole.
A vortex turns out to generate an event horizon for light,
a radius of no return, beyond which light falls inevitably towards
the vortex singularity.
Similar to a star that turns into a black hole when the gravitational
Schwarzschild radius exceeds the star's size, 
a vortex appears as a black hole
when the optical Schwarzschild radius exceeds the radius of the core
(the size of the ``eye of the hurricane'').

Optical effects of moving media have been known for a long time.
In 1818 Fresnel \cite{Fresnel} concluded correctly from an ether
theory that a moving medium will drag light.
Fizeau \cite{Fizeau} observed Fresnel's drag effect in 1851.
In 1895 Lorentz \cite{Lorentz} derived an additional drag component
that is due to optical dispersion (the frequency--dependence of 
the refractive index).
Zeeman \cite{Zeeman} was able to verify experimentally Lorentz' effect.
In 1923 Gordon \cite{Gordon} formulated the electromagnetism
in dispersionless media in terms of an effective gravitational field
(an effective non--Euclidean metric).
Let us develop a theory of light propagation in highly dispersive
and transparent media in the spirit of Gordon's 
{\it Lichtfortpflanzung nach der Relativit\"atstheorie} \cite{Gordon}.

{\it The model.---}
Imagine that a dielectric consists of small volume elements. 
Each element is sufficiently small such that the refractive index $n$
and the medium velocity ${\bf u}$ do not vary significantly,
but each volume element is large enough to sustain several optical
oscillations. 
We thus assume that the properties of the dielectric do not vary
substantially over the effective optical wave length in the medium. 
In this case the propagation of light in each medium element does
not depend on the polarization, 
and we can describe light waves by the scalar dispersion relation
\begin{equation}
\label{dm}
k'^2 - \frac{\omega'^2}{c^2} - \chi(\omega')\,\frac{\omega'^2}{c^2} = 0
\,\,.
\end{equation}
Here ${\bf k'}$ denotes the local wave vector,
$\omega'$ is the local optical frequency,
and we use primes to distinguish quantities in locally co--moving 
medium frames. 
Let us specify the susceptibility $\chi(\omega')$.

Electromagnetically induced transparency \cite{EIT}
has been applied to create dielectrics with extraordinary low 
group velocity \cite{Lene}.
Here a coherent electromagnetic wave drives the atoms of the medium
into a quantum--superposition state such that a probe wave 
can travel through the dielectric that would otherwise be completely opaque.
Under ideal circumstances the probe experiences at a certain frequency 
$\omega_0$ a vanishing susceptibility $\chi$
and a real (and extremely low) group velocity without 
group--velocity dispersion \cite{Harris}.
We thus assume that in the spectral vicinity of $\omega_0$
the susceptibility is, up to terms of third order in
$\omega'-\omega_0$,
\begin{equation}
\label{chi}
\chi(\omega') = \frac{2\alpha}{\omega_0}\,(\omega'-\omega_0) + 
{\rm O} \left((\omega'-\omega_0)^3\right)
\,\,.
\end{equation}
We obtain from Eqs.\ (\ref{dm}) and (\ref{chi}) the
group velocity
\begin{equation}
\label{vg}
v_g \equiv \left. \frac{\partial \omega'}{\partial k'}
\right|_{\omega_0} =
\left(\left. \frac{\partial k'}{\partial \omega'}
\right|_{\omega_0}\right)^{-1} =
\frac{c}{\alpha + 1}
\,\,.
\end{equation}
For having a definite geometry in mind, we imagine that the medium 
flow is perpendicular to one axis in space.
The driving wave shall run in the direction of this axis,
i.e. orthogonally to the motion of the medium.
This arrangement has the advantage that the atoms of the medium are
not sensitive to the first--order Doppler effect of the drive
(and higher--order effects turn out to be irrelevant for our purpose).
A monochromatic probe beam of frequency $\omega_0$ shall propagate
orthogonally to the driving beam, i.e. in the plane where the medium moves.
Consequently, the probe experiences the full range of Doppler--detuning 
of the susceptibility (\ref{chi}) at the sharp resonance $\omega_0$,
while still propagating in a transparent medium.

{\it The metric.---}
How does the moving medium appear to the probe?
Let us transform the dispersion relation (\ref{dm}) in locally 
co--moving  medium frames to the laboratory frame.
We notice that $k'^2-\omega'^2/c^2$ is a Lorentz scalar,
and obtain 
\begin{equation}
\label{dr}
k^2 - \frac{\omega_0^2}{c^2} - \chi(\omega')\,\frac{\omega'^2}{c^2} = 0
\,\,,\quad
\omega' = 
\frac{\omega_0 - {\bf u} \cdot {\bf k}}{\sqrt{1-\frac{u^2}{c^2}}}
\,\,,
\end{equation}
where ${\bf u}$ denotes the velocity field of the medium.
Note that the local Lorentz transformations from the medium frames 
to the laboratory frame mix the components of the electromagnetic
field--strength tensor $F_{\mu \nu}'$.
However, since the dispersion relation (\ref{dm}) in the medium
is valid for all components of $F_{\mu \nu}'$,
the light propagation in the laboratory frame is 
polarization--independent \cite{LP}.
Relativistic effects of light in slowly moving media
diminish with increasing order in $u/c$.
Therefore, we expand the dispersion relation (\ref{dr}) to second
order in $u/c$, use the susceptibility (\ref{chi}), and arrive at
a result that we can formulate in the spirit of Gordon's
geometric theory \cite{Gordon,LP}.
We introduce the covariant wave vector,
\begin{equation}
k_\nu =
\left( \frac{\omega_0}{c}, -{\bf k}\right) 
\,\,,
\end{equation}
and, adopting Einstein's summation convention,
obtain the dispersion relation
\begin{equation}
\label{d2}
g^{\mu\nu} k_\mu k_\nu = 0
\end{equation}
with
\begin{equation}
\label{metric}
g^{\mu\nu} =
\left(
\begin{array}{cc}
1+\alpha {\displaystyle \frac{u^2}{c^2}} & 
\alpha {\displaystyle \frac{\bf u}{c}} \\
\alpha {\displaystyle \frac{\bf u}{c}} & 
-{\bf 1} + 4\alpha {\displaystyle \frac{{\bf u} \otimes {\bf u}}{c^2}}
\end{array}
\right)
\,\,.
\end{equation}
The symbol $\otimes$ denotes the three--dimensional tensor product.
We regard $g^{\mu\nu}$ as the contravariant metric tensor of the 
moving medium, whereas the inverse of $g^{\mu\nu}$ is the 
covariant tensor $g_{\mu\nu}$.
Light rays turn out \cite{Gordon,LP} to be zero--geodesic lines of
\begin{equation}
ds^2 = g_{\mu\nu} dx^\mu dx^\nu 
\,\,,\quad
dx^\mu = \left(c\,dt, d{\bf x} \right)
\,\,.
\end{equation}
The moving dielectric appears as a curved space--time,
i.e. as an effective gravitational field, 
to light that travels inside.

Let us introduce the contravariant wave vector $k^\mu$ with respect
to the metric $g^{\mu\nu}$ of the medium,
\begin{equation}
\label{contrak}
k^\mu \equiv g^{\mu\nu} k_\nu
\,\,.
\end{equation}
One can show \cite{LP} that this four--vector points into the
direction of light propagation,
\begin{equation}
k^\mu = \frac{k^0}{c}\,\frac{dx^\mu}{dt}
\,\,.
\end{equation}
In other words, $k^\mu$ is proportional to the velocity vector of light, 
i.e.\ $k^\mu$ appears as a kinetic momentum, 
whereas the covariant wave vector $k_\nu$ is the canonical momentum 
of the light ray.

{\it Optical Aharonov--Bohm effect.---}
The distinction between canonical and kinetic momentum is as vital
to the physics of charged particles in magnetic fields as the
distinction of co- and contravariant vectors is to general
relativity.
The two areas are related.
In fact, we obtain from the definition (\ref{contrak}) and the metric
(\ref{metric}) to lowest order in $u/c$
\begin{equation}
k^0 = \frac{\omega_0}{c} - \alpha \frac{{\bf u} \cdot {\bf k}}{c} 
\,\,,\quad
k^i = {\bf k} + \alpha \frac{\omega_0}{c}\,\frac{{\bf u}}{c}
\,\,.
\end{equation}
We apply the dispersion relation (\ref{d2}) and get,
up to second--order terms,
\begin{equation}
\label{k3}
\sum_{i=1}^3 (k^i)^2 = \frac{\omega_0^2}{c^2}
\,\,.
\end{equation}
In geometrical optics \cite{LP}
one can translate a dispersion relation into 
a wave equation for the complex positive--frequency component of 
the field--strength tensor $F_{\mu\nu}$ 
by substituting $-i\nabla$ for ${\bf k}$.
In particular, we obtain from the relation (\ref{k3})
\begin{equation}
\label{wave}
\left(-i\nabla + \alpha\,\frac{\omega_0}{c^2}\,{\bf u}\right)^2
F_{\mu\nu} = \frac{\omega_0^2}{c^2} F_{\mu\nu}
\,\,.
\end{equation}
This is precisely the non--relativistic Schr\"odinger equation
of a charged matter wave in a magnetic field.
Consequently, light in a slowly moving dispersive medium behaves like
an electron wave where the medium velocity {\bf u} plays the role
of the vector potential \cite{OpticalAB}.

Aharonov and Bohm discovered \cite{AB}
that under certain circumstances a charged matter wave attains an
observable phase shift without experiencing a force.
In particular, a thin solenoid produces a vanishing magnetic field
outside the coil, and hence generates no force, and yet,
matter waves that enclose the solenoid experience a noticeable 
phase shift due to a vortex of the vector potential.
Consequently, in the case of light in moving media,
a vortex flow will not bend light in first order,
but the vortex will imprint a phase shift onto the incident light
\cite{OtherAB,Roentgen}.
In cylindrical coordinates a vortex with vorticity $2\pi{\cal W}$
has the velocity profile
\begin{equation}
\label{vortex}
{\bf u} = \frac{\cal W}{r}\,{\bf e}_\varphi
\,\,.
\end{equation}
We compare the wave equation (\ref{wave}) with the Schr\"odinger 
equation of Aharonov and Bohm \cite{AB}, and read off the phase
shift
\begin{equation}
\varphi_{_{AB}} = 2\pi \nu_{_{AB}}
\,\,,\quad
\nu_{_{AB}} = \alpha\,\frac{\omega_0}{c} \frac{\cal W}{c} =
\frac{\omega_0}{c} \frac{\cal W}{v_g}
\end{equation}
in the limit of a low group velocity $v_g$.
Electromagnetically induced transparency has made it possible
to reduce $v_g$ to $17{\rm m}/{\rm s}$ \cite{Lene}.
In this case the optical Aharonov--Bohm effect is sensitive enough
to detect a single quantum vortex with
\begin{equation}
\label{quantumvortex}
{\cal W} = \frac{\hbar}{m}
\,\,.
\end{equation}
Indeed, we obtain for a frequency $\omega_0$ of 
$3\times 10^{15}{\rm s}^{-1}$
and for sodium with a rest mass $m$ of about 23 proton masses a 
$\varphi_{_{AB}}$ of $10^{-2}$.
This phase shift between waves that pass the vortex from different
sides can be made visible via phase--contrast microscopy \cite{Andrews}.
The optical Aharonov--Bohm effect explores the
long--ranging topological nature of a quantum vortex,
similar to the vortex detection \cite{BWTD}
using two interfering condensates.
In contrast to this technique, 
the optical effect may allow {\it in--situ} observations of vortices.

{\it Optical black hole.---}
A classical vortex generates a strongly falling pressure near the
vortex core.
A tornado, for example, attracts with ease substantial ``test
particles'' and tears them apart. 
Can a vortex attract light?
What happens near the core where the first--order 
Aharonov--Bohm theory is destined to fail?
Let us study the light propagation using the Hamilton--Jacobi method
\cite{LL1}.
Here the covariant wave vector $k_\nu$ is the negative four--gradient
of the eikonal $S$, or,
\begin{equation}
\omega_0 = -\frac{\partial S}{\partial t}
\,\,,\quad
{\bf k} = \nabla S
\,\,.
\end{equation}
We interpret the dispersion relation (\ref{d2}) as the
Hamilton--Jacobi equation for light rays.
In the case of the vortex flow (\ref{vortex})
we find in cylindrical coordinates
\begin{eqnarray}
\label{eikonal}
S & = & \frac{\omega_0}{c} \left[
-ct + l\varphi + R(r) \right]
\,\,,
\\
\left(\frac{dR}{dr}\right)^2 
& = &
1 - \frac{l^2}{r^2} + 
\frac{\alpha}{r^2}\left(-2\frac{{\cal W}l}{c} + \frac{{\cal W}^2}{c^2} +
4\,\frac{{\cal W}^2 l^2}{c^2 r^2} \right)
\,\,.
\nonumber
\end{eqnarray}
The eikonal (\ref{eikonal}) characterizes a set of rays
with common frequency $\omega_0$ and angular momentum $l (\omega_0/c)$
that are incident perpendicular to the vortex line.
Note that near the origin the modulus of the wave vector grows 
at least as rapidly as the flow velocity.
The ratio $|\nabla S|/u$ approaches here 
$2\sqrt{\alpha}\,(l/r)(\omega_0/c^2)$
for $l \neq 0$ and $\sqrt{\alpha}\, (\omega_0/c^2)$ for $l=0$.
Consequently, even in the vicinity of the vortex core,
geometrical optics is well justified to describe 
the propagation of light.

How close can light come to the core and still manage to escape?
Let us analyze the turning points $r_0$ of the radial motion
where $(dR/dr)^2$ vanishes.
For each value of $l$ we obtain two points,
an outer $(+)$ and an inner turning point $(-)$,
\begin{equation}
\label{r02}
r_0^2 = \frac{1}{2c^2} 
\left(w_0^2 \pm \sqrt{w_0^4-16\alpha c^2 l^2{\cal W}^2} \right)
\,\,,
\end{equation}
\begin{equation}
\label{w02def}
w_0^2 = (\alpha+1) c^2l^2 - \alpha (cl-{\cal W})^2
\,\,,
\end{equation}
provided that the argument of the square root in Eq.\ (\ref{r02})
is non--negative.
Otherwise real turning points do not exist,
and the incident light is doomed to fall towards the vortex core.
At two critical angular momenta $l_\pm$ the inner and the outer
turning points coincide at $r_\pm \equiv r_0$.
In this case we get from Eq.\ (\ref{r02}) the relation
\begin{equation}
\label{w02}
w_0^2 = \pm 4 \sqrt{\alpha}\, {\cal W}\, c l_\pm
\end{equation}
with, assuming a positive vorticity $2\pi{\cal W}$,
the plus sign for positive $l_+$ and the minus sign for negative $l_-$,
because $w_0^2$ is non--negative.
Light rays with angular momenta inside the interval $(l_-,l_+)$
have no turning points.
Consequently,
the critical angular momenta $l_\pm$ mark the transition from the fall
into the core and a chance to escape.
We solve Eqs.\ (\ref{w02def}) and (\ref{w02}) for $l_\pm$, obtain
\begin{equation}
l_\pm = \frac{{\cal W}}{c}\sqrt{\alpha} \left[
-\sqrt{\alpha} \pm 2 \pm \sqrt{(-\sqrt{\alpha} \pm 2)^2 + 1} \right]
\,\,,
\end{equation}
and get in the limit of low group velocities $v_g$
when $\alpha \sim c/v_g \gg 1$,
\begin{equation}
\label{lpm}
l_- = -2\frac{{\cal W}}{v_g}
\,\,,\quad
l_+ = \frac{{\cal W}}{2c}
\,\,.
\end{equation}
Finally, we obtain from Eqs.\ (\ref{r02}), (\ref{w02}) and (\ref{lpm})
the corresponding critical radii $r_\pm = r_0$,
where turning points cease to exist,
\begin{equation}
r_- = 2\frac{{\cal W}}{c} \left(\frac{c}{v_g}\right)^{3/4}
\,\,,\quad
r_+ = \frac{{\cal W}}{c} \left(\frac{c}{v_g}\right)^{1/4}
\,\,.
\end{equation}
Regardless which path the light is following, as soon as a light ray
comes closer than $r_+$ to the vortex core, the light faces no other
choice than to fall towards the singularity.
The optical Schwarzschild radius $r_+$ determines a point
of no return
(in contrast to trajectories in other singular potentials
\cite{atomsandwire} 
where escaping particles may come arbitrarily close to the singularity).
The larger critical radius $r_-$ is a weak Schwarzschild radius
where light rays with positive angular momenta can escape but those
with negative $l$ are trapped.
Light rays with positive angular momentum have the advantage of
traveling with the flow, whereas those with negative $l$ swim against 
the current, and are efficiently deaccelerated and finally captured.

One might object that the vortex flow (\ref{vortex}) of our model
will allow medium velocities that exceed $c$ near the origin.
Note, however, that the flow velocities $u_\pm$ at the two
Schwarzschild radii are well below $c$,
\begin{equation}
u_- = \frac{c}{2} \left(\frac{v_g}{c}\right)^{3/4}
\,\,,\quad
u_+ = c \left(\frac{v_g}{c}\right)^{1/4}
\,\,.
\end{equation}
Long before the vortex (\ref{vortex}) becomes superluminal,
the falling pressure will produce a hole in the vortex core
(the ``eye of the hurricane'').
The vortex appears as an optical black hole
if the core radius is smaller than the Schwarzschild radius.
Suppose that one could reduce the group velocity of light further to
$1 {\rm cm}/{\rm s}$.
In this case the velocity $u_+$ at the hard Schwarzschild radius $r_+$
reaches $7\times 10^5  {\rm m}/{\rm s}$ 
and the flow at the weak Schwarzschild radius $r_-$ is 
$2{\rm m}/{\rm s}$.
The creation of a hard black hole seems to be unrealistic with 
present technology.
However, a weak black hole could be made.
For example, one could utilize the torque of Gauss--Laguerre beams
\cite{Babiker} to create a classical vortex of a rapidly rotating
gas of alkali atoms \cite{experiments}.
Especially appealing would be a quantum black hole with a 
single quantum vortex (\ref{quantumvortex}) 
as the center of attraction.
In alkali Bose--Einstein condensates \cite{BEC},
the core radius is roughly given by the healing length 
$(8\pi\rho a)^{-1/2}$ with $\rho$ being the density and $a$
the scattering length.
For a sodium condensate ($\rho = 5\times 10^{18}{\rm m}^{-3}$
\cite{Lene} and $a = 2.75\times 10^{-9}{\rm m}$)
we obtain a healing length of about $2\times 10^{-6}{\rm m}$
that significantly exceeds the Schwarzschild radius $r_-$ of about 
$10^{-9}{\rm m}$.
However, one could employ other alkali isotopes and/or utilize Feshbach
resonances \cite{Inouye} to increase the scattering length and,
consequently, to reduce the size of vortex cores.

{\it Summary.---}
A moving dielectric medium acts as an effective gravitational field on
light \cite{Gordon,LP}.
One could employ media with extremely low group velocities
\cite{Lene} to create dielectric analogs of gravitational effects that
usually belong to the realm of astronomy.
In particular, a vortex can create a long--ranging Aharonov--Bohm 
effect on incident light \cite{OpticalAB,AB,OtherAB,Roentgen} 
and, on shorter ranges, can behave like a black hole \cite{SonicBH}.

We are grateful to
M.~V. Berry,
J.~H. Hannay,
S. Klein,
W. Schleich,
and
S. Stenholm
for helpful discussions.
U.~L. thanks the Alexander von Humboldt Foundation
and the G\"oran Gustafsson Stiftelse for support.
P.~P. was partially supported by the 
research consortium {\it Quantum Gases} of the
Deutsche Forschungsgemeinschaft.


\end{document}